# A Workflow for Utilizing OpenFOAM Data Structure in Physics-Informed Deep Learning Training

Yijin Mao and Yuwen Zhang[1]

*Department of Mechanical and Aerospace Engineering, University of Missouri,*

*Columbia, MO, 65211*

**Abstract:**

This study presents a novel methodology for integrating physics-informed loss functions into deep learning models using OpenFOAM's comprehensive data structures. Leveraging the robust and flexible capabilities of OpenFOAM's data structure for handling complex geometries and boundary conditions, it is demonstrated how to construct detailed loss functions that accurately embed physics constraints and potentially enhance the training and performance of neural networks in handling industrial-level complicated geometry for computational fluid dynamics (CFD) simulations. The present work primarily focuses on the 1D Burger equation to showcase the detailed procedure of constructing initial loss, boundary loss, and residual loss. While the computational geometry employed here is relatively simple, the procedure is sufficiently general to illustrate its applicability to more complex computational domains. The results show the trained operator former (OFormer) neural network can successfully predict the simulation results subject to the OpenFOAM's data structure composed loss. This framework potentially opens new avenues for using deep learning to tackle complex industrial simulation challenges, promising significant advancements in the accuracy and practicality of CFD applications.

## 1 Introduction:

In the evolving landscape of computational fluid dynamics (CFD), the integration of machine learning techniques, particularly deep learning, offers promising enhancements in simulation accuracy and computational efficiency. Physics-informed neural networks (PINNs) [1-5] have emerged as a powerful tool, enabling the incorporation of physical laws directly into the architecture of neural models. This integration ensures that simulations not only conform to but are guided by the underlying physical principles governing real-world phenomena, adding more interpretability to the model behavior. However, the effectiveness of PINNs in solving real-world CFD problems critically depends on the precision with which data structures handle the geometrical and physical complexities of the simulated domain. OpenFOAM [6], an open-source CFD platform renowned for its robust and flexible handling of complex geometries, provides an ideal framework for this task. Its ability to manage detailed and intricate mesh geometries makes it particularly suitable for training neural networks that require high fidelity to physical laws and boundary conditions.

This paper presents a technical exploration of utilizing OpenFOAM's data structures to construct detailed, physics-informed loss functions for neural network training. This work focuses on the methodological integration of OpenFOAM's mesh-based data into the training process, enhancing the neural network's ability to learn from and adhere to the dynamics dictated by CFD simulations. Through the specific case of the 1D Burger equation [7], a prototypical problem in fluid dynamics, it is demonstrated how OpenFOAM's data can be systematically utilized to embed into the loss functions of deep learning models.

---

[1]Corresponding Author. Email: zhangyu@missouri.edu

## 2 Methodology

Following OpenFOAM's standard discretization procedure, this study utilizes the finite volume method (FVM) to discretize the 1D Burger Equation, a standard model in fluid dynamics that exemplifies non-linear advection processes.

$$\frac{\partial \mathbf{U}}{\partial t} + \nabla \mathbf{U}\mathbf{U} = \nabla^2 \nu \mathbf{U} \tag{1}$$

where $\mathbf{U}$ represents the velocity field and $\nu$ the kinematic viscosity. It is worth mentioning that the presented methodology highlights the integration of OpenFOAM's mesh data structures into the construction of physics-informed loss functions for neural networks. This work is intended to demonstrate the potential of incorporating complex geometry data into the loss function formulation. While the computational geometry employed here is relatively simple, the procedure is sufficiently general to illustrate its applicability to more complex computational domains [8, 9].

### 2.1 Problem Setup

#### 2.1.1 **Finite Volume Discretization of the Burger Equation**

The Burger Equation is discretized using the Gaussian theorem, which integrates fluxes across control volumes, and a first-order Euler implicit temporal scheme. The spatial discretization involves linear interpolation to express convection and diffusion terms between computational cells, calculated as follows:

$$\frac{\mathbf{U}_i^{t+1} - \mathbf{U}_i^t}{\Delta t} + \frac{\sum_f ((\omega_{f,o}\mathbf{U}_o + \omega_{f,n}\mathbf{U}_n) \cdot \mathbf{S}_f)(\omega_{f,o}\mathbf{U}_o + \omega_{f,n}\mathbf{U}_n)}{\Delta V} \\ = \frac{\sum_f (\omega_{f,o}(\nu\nabla\mathbf{U})_o + \omega_{f,n}(\nu\nabla\mathbf{U})_n) \cdot \mathbf{S}_f}{\Delta V} \tag{2}$$

where $\mathbf{S}_f$ represents the surface vector of the control volume faces and subscripts *o* and *n* refer to the *owner* and *neighbour* cells of the associated cell face, respectively. The weighting coefficients $\omega_o$ and $\omega_n$, which are crucial for achieving accurate interpolation between cells, are defined based on the distances between cell centers and the face centroid, as depicted in Figure 1.

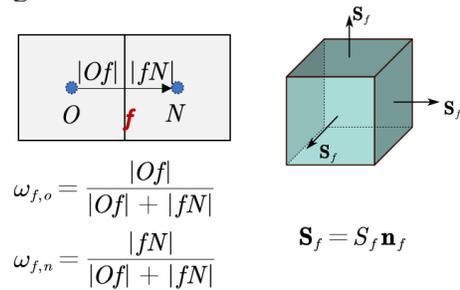

Figure 1: The depicted diagram illustrates the distance weighting coefficients and cell face area vector of the control volume.

It is worth mentioning that the gradient of velocity, in Eq. (2), can be found through the Gaussian theorem as well via using the same field and geometric information as other discretization procedures, as shown in Eq. (3).

$$\nabla \mathbf{U} = \frac{\sum_f \mathbf{U}_f \otimes \mathbf{S}_f}{\Delta V} \qquad (3)$$

### 2.1.2 Boundary Condition Handling

In the context of the Burger Equation, the boundary conditions are defined to ensure physics sound and unique solutions in simulations. Fixed boundary conditions are applied at both ends of the domain, setting the velocity $\mathbf{U}$ to 0 m/s.

For all other boundaries, zero-gradient conditions are imposed, ensuring no change in velocity across these boundaries, and address the 1D simulation problem (Typically, empty type could be applied, however, here zero gradient is chosen for a full 3D description of the problem). The boundary gradient is calculated using the normal component as in Eq. (4),

$$\nabla_n \mathbf{U} = \nabla \mathbf{U} \cdot \mathbf{n}_f \qquad (4)$$

where $\mathbf{n}_f$ represents the outward normal vector to the boundary face, ensuring that the boundary conditions are accurately represented in the numerical model.

### 2.1.3 Initial Conditions

The initial condition for the velocity field across the domain is defined by a sinusoidal function, as given in Eq. (5),

$$\mathbf{U}^0(\mathbf{x}) = -\sin(\pi \mathbf{x}) \qquad (5)$$

It sets up a wave-like velocity profile at the start of the simulation. This initial state is significant as it establishes the initial dynamics from which the fluid behavior evolves over time.

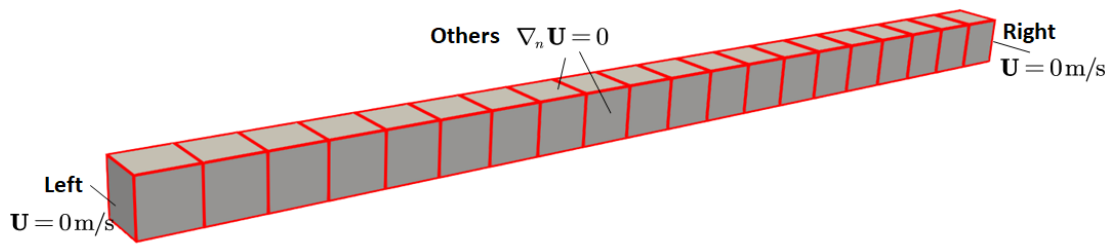

Figure 2: Illustration of the boundary conditions and computational domain (For clear illustration, here only 20 cells are used. Actually training setup uses 200 cells.)

### 2.2 Incorporation into the Loss Function

As presented in the Eqs. (2)-(4), to successfully embed the governing equations, along with the boundary and initial conditions into the loss function, it is essential to utilize mesh-related parameters such as distance weights ($\omega$), cell volume ($\Delta V$), cell face area ($\mathbf{S}_f$), cell coordinates (including cell center and boundary face centers), and field-related parameters like the velocity field. These details are vital for constructing a loss function that accurately reflects both the physics of the problem and the specifics of the computational domain. This comprehensive integration ensures that the trained model adheres closely to the expected physical behaviors dictated by both the fluid dynamics, the boundary, and initial conditions.

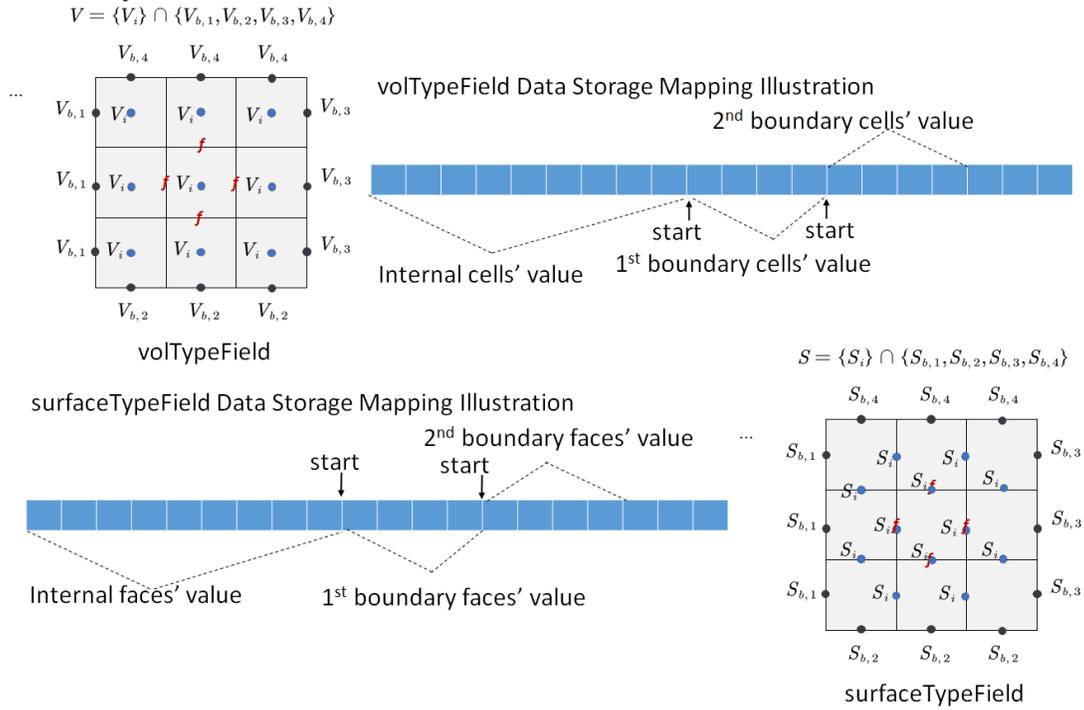

Figure 3: Data Storage and Mapping for volTypeField and surfaceTypeField. The top panel illustrates the volTypeField data structure, showing the integration of cell values and boundary face values. The bottom panel displays the surfaceTypeField data structure, showing how values are assigned to internal and boundary faces. It should be noted that the 2D grids illustrate the data storage in OpenFOAM while the 1D bar is for data storage in the present work.

### 2.2.1 Mesh and Field Data Structures in OpenFOAM

OpenFOAM's data structures, volTypeField and surfaceTypeField (here Type could be scalar, vector, tensor, etc.,), are fundamental in facilitating the application of the finite volume method, ensuring precise spatial discretization and boundary treatment, as shown in Figure 3. A volTypeField is composed of a list of internal cell values and a list of list of boundary face values. It integrates values from both the cell interiors and their boundaries, thereby accommodating changes across the mesh geometry effectively. In the present case, this field type is critical for capturing the velocity at each cell center point, essential for calculating fluxes and other properties accurately. The storage mapping of volTypeField illustrates how data from adjacent cells and faces are compiled, offering a clear picture of how velocities at boundaries and internal points are managed (This mapping is for the present work's data storage. OpenFOAM uses internalField and boundaryFields to manage these data.) This structured arrangement ensures that each cell's contribution to the overall field is correctly accounted for and easily accessible for computations. In contrast, the surfaceTypeField, which is also composed of of list of internal face values and a list of list of boundary face values, is designed to specifically manage data associated with the faces of cells. It plays a vital role in representing interfacial properties such as fluxes across faces, which are crucial for boundary and internal face calculations. Particularly, it is also very important for geometry-relevant representation parameters, such as distance weighing coefficient, surface area, and surface normal direction. The mapping of surfaceTypeField shown in the figure demonstrates the distinction between internal and boundary faces, providing a systematic way to access and manipulate face-specific data efficiently. In Figure 3, the diagrams also clarify the data storage and mapping for both volTypeField and surfaceTypeField. The first N items, where N corresponds to the total number of cells, represent cell center values for volTypeField. The subsequent entries in the array are designated for values associated with different boundary patches (each patch knows the start and end face index). This convention also applies to surfaceTypeField field.

### 2.2.2 Face-Cell Indexing and Data Structure Utilization in OpenFOAM

In OpenFOAM, the mesh data structure, essential for the management of both geometric and physical processes, is effectively organized through key face-cell indexing. This system ensures efficient access to geometric and dynamic properties such as distance weights, cell volume, face area, cell center coordinates, and face normal vectors necessary for computations. The mesh topology in OpenFOAM is defined by the *owner* and *neighbour* files, as shown in Figure 4, which are fundamental in establishing the relationships between faces and cells. The *owner* file contains indices of cells that own each face, crucial for defining the face-cell relationship necessary for applying boundary conditions and managing flow dynamics. Conversely, the *neighbour* file lists the neighboring cells for each internal face, supporting the calculation of gradients and other differential properties across cells. These indices are pivotal in defining the structure of the computational domain, allowing precise control and manipulation of data for simulation tasks.

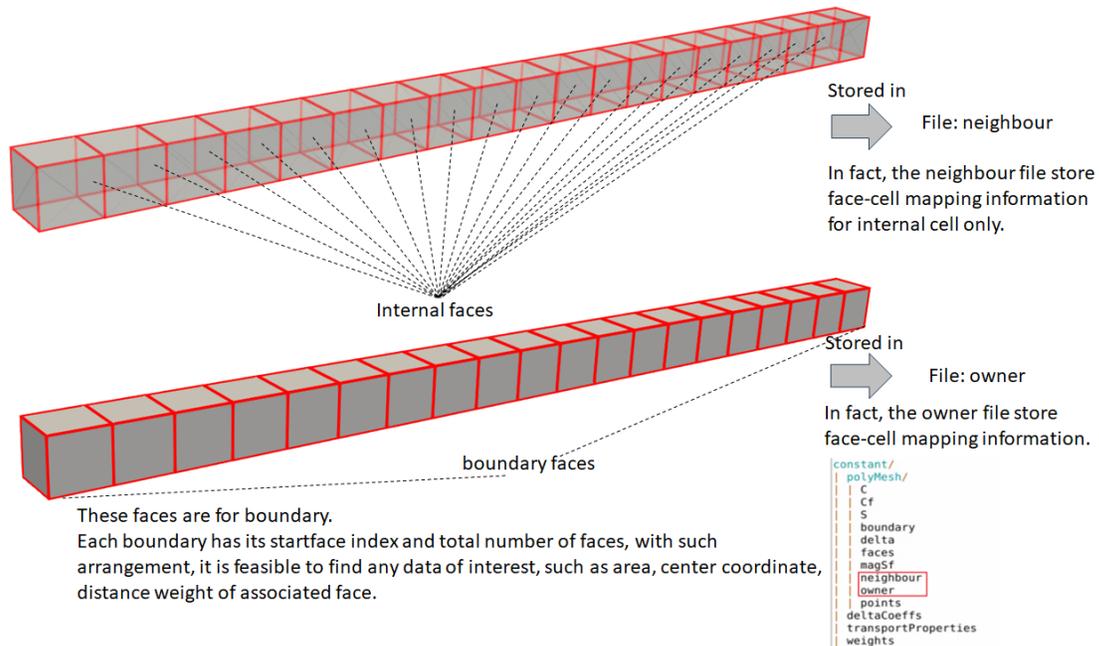

Figure 4: It details the organization of internal and boundary faces within the OpenFOAM mesh structure. It shows the storage of face-cell mapping information in the '*owner*' and '*neighbour*' files, which are essential for defining relationships within the mesh. The arrangement facilitates efficient access to critical data such as face areas, center coordinates, and distance weights, crucial for computational fluid dynamics simulations

### 2.2.3 Utilizing the Face-Cell Indexing for Simulations

Integrating physics laws and boundary constraints is accomplished through a systematic process of iterating over all faces, as detailed in the flow chart in Figure 5. This involves looping through all the faces listed in the *owner* file to apply governing equations and handle boundary treatments effectively. During each iteration, the relationships defined in the *owner* and *neighbour* files facilitate access to and modification of properties of cells that share a common face. This process ensures that both residual and boundary treatments are correctly and efficiently applied, maintaining the integrity and accuracy of the simulation.

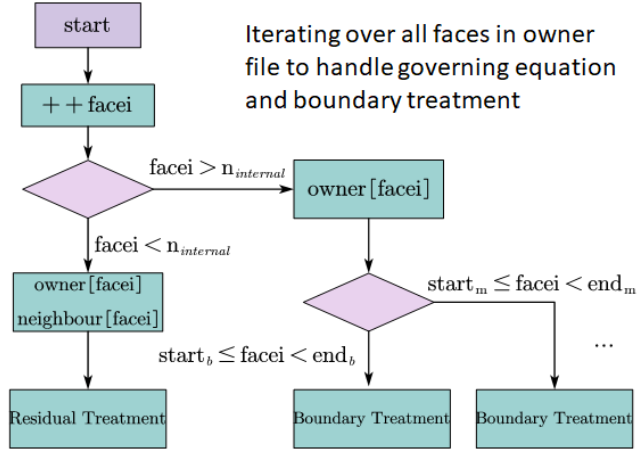

Figure 5: The flowchart demonstrates the procedural steps for applying governing equations and boundary treatments by iterating over the mesh faces (in the *owner* file), ensuring comprehensive and precise implementation of physical law constraints.

As shown in Figure 5, by iterating over these relationships, this data structure ensures that all physical laws and boundary conditions are consistently imposed across the computational mesh. Here total internal faces is the number of faces stored in *neighbour* file, the start face index for each boundary can be retrieved from a file named *boundary* (see at bottom-right of Figure 4). This leverages the detailed mesh data structure for accurate and effective computational fluid dynamics simulations.

### 2.2.4 Constructing Loss for the Burger Equation

The construction of the loss function for the Burger Equation involves three primary components: residual loss, boundary condition loss, and initial condition loss. Each of these components targets a specific aspect of the simulation, ensuring that the trained model accurately reflects both the dynamics of the fluid and the conditions at the boundaries of the computational domain.

#### 2.2.4.1 Residual Loss, $L_r$

The residual loss (via mean squared error-MSE [10]) is formulated to ensure that the solution adheres to the discretized Burger Equation across all cells in the computational domain. It is defined as:

$$L_r = \frac{1}{N_{cell}} \sqrt{\sum_{N_{cell}} \left[ \frac{\mathbf{U}_i^{t+1} - \mathbf{U}_i^t}{\Delta t} + \frac{\sum_f ((\omega_{f,o}\mathbf{U}_o + \omega_{f,n}\mathbf{U}_n) \cdot \mathbf{S}_f)(\omega_{f,o}\mathbf{U}_o + \omega_{f,n}\mathbf{U}_n)}{\Delta V} - \frac{\sum_f (\omega_{f,o}(\nu\nabla\mathbf{U})_o + \omega_{f,n}(\nu\nabla\mathbf{U})_n) \cdot \mathbf{S}_f}{\Delta V} \right]^2} \quad (5)$$

where $N_{cells}$ refers to the total number of cells. This component penalizes the deviation of the numerical solution from the conservation laws governing the fluid flow, integrating over all cells to capture the global behavior of the system.

#### 2.2.4.2 Boundary condition loss, $L_b$

Boundary condition loss (via mean squared error-MSE) ensures that the model respects the boundary conditions specified for the problem. This loss is particularly crucial for maintaining physical realism at the domain edges:

$$L_b = \frac{1}{N_{b,faces}} \sqrt{\left[ \sum_f^{N_{left,faces}} \mathbf{U}_f^2 + \sum_f^{N_{right,faces}} \mathbf{U}_f^2 + \sum_f^{N_{others,faces}} [(\nabla \mathbf{U})_f \cdot \mathbf{n}_f]^2 \right]} \tag{6}$$

where $N_{b,faces}$ refers to the total number of boundary faces and $\mathbf{n}_f$ is the outward normal at the face.

#### 2.2.4.3 Initial condition loss, $L_0$

The initial condition loss (via mean squared error-MSE) compares the model's predicted initial state to the actual physical initial conditions provided:

$$L_0 = \frac{1}{N_{cells}} \sqrt{\sum_i (\mathbf{U}_i(t_0) - \mathbf{U}_i^0)^2} \tag{7}$$

where $\mathbf{U}_i^0$ is the initial velocity field specified for the simulation. This loss helps align the model's predictions with the known starting conditions of the fluid.

Since the adopted neural network model is designed for predicting time-dependent output, as illustrated in Figure 6, the dynamic evolution of the fluid's velocity field from the initial time $t_0$ through subsequent time steps $t$ and $t+1$, until the last timestep, substituting the prediction time sequence data into the loss definitions (Eqns. (5)-(7)), the complete loss function, embedded residual, boundary, and initial constraints, can be fully constructed.

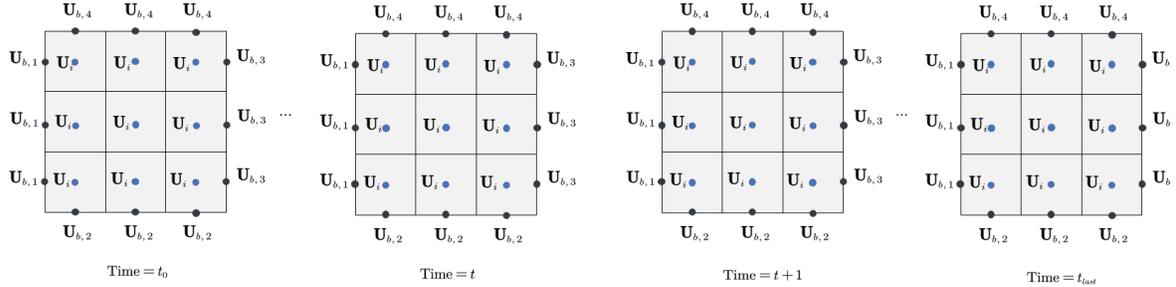

Figure 6: Predicted time sequence subject to volTypeField data structure

The entire training process is conducted within the PyTorch framework [11]. The neural network model employed is based on the transformer architecture of PDEs operator learnining [12]. Initial hyperparameter tuning is facilitated by the use of a OneCycleLR [13] scheduler (2500 steps), which helps determine the optimal initial learning rate of $1.68 \times 10^{-4}$. A multiple-step scheduler is used for further optimization of 12500 steps. The AdamW optimizer [14] is utilized to navigate the parameter space, aiming to minimize the total loss. The computational domain spans from -1 to 1, with the kinematic viscosity set at 1.0. The model discretizes the domain into 200 uniform cells to ensure a detailed resolution of the fluid's behavior. The CFD simulation results, essential for training validation, are generated using a customized OpenFOAM solver. Interested readers are invited to contact the authors via email to obtain a copy of the complete code.

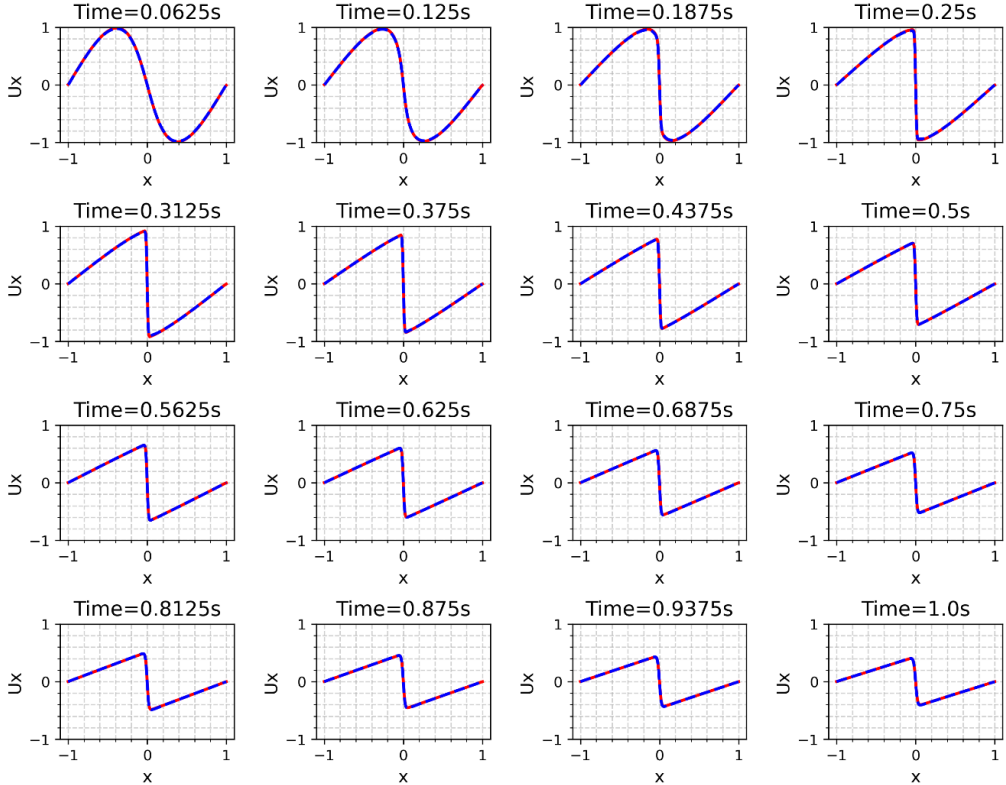

Figure 7: Comparison of predicted velocity distribution across the domain with CFD solver solution (Red: predicted; Blue: CFD simulation data).

## 3 Result and Discussions

The results presented in Figure 7 demonstrate the temporal progression of the velocity field $\mathbf{U}_x$ predicted by the trained neural network model against that of the CFD solver. These figures show the velocity distribution at various time steps (processed with vtk [15, 16]), ranging from t=0.0625s to t=1.0s, across the computational domain of [−1,1]. The sequence of plots effectively captures the dynamic behavior of the velocity field as governed by the Burger Equation. As can be seen, at earlier time steps, the model predicts a sinusoidal velocity distribution, which gradually transitions into sharper profiles as time progresses. This change is indicative of the nonlinear advection processes characteristic of the Burger Equation and highlights the model's capability to adapt to the evolving dynamics of the system. The predicted results are compared against those obtained from a customized OpenFOAM solver, which serves as a benchmark for evaluating the accuracy of the neural network. The close alignment between the neural network predictions and the CFD solver results validates the effectiveness of the physics-informed loss functions in guiding the neural network toward physically plausible solutions. This alignment also demonstrates the model's ability to generalize well across different time steps and initial conditions via employing the OpenFOAM data structure.

Additionally, spanning time intervals from 0.1s to 1.0s and via mapping prediction back to the OpenFOAM data structure, ten velocity contour plots are provided to visually demonstrate the spatial distribution and evolution of the velocity field over the simulation period, as shown in Figure 8.

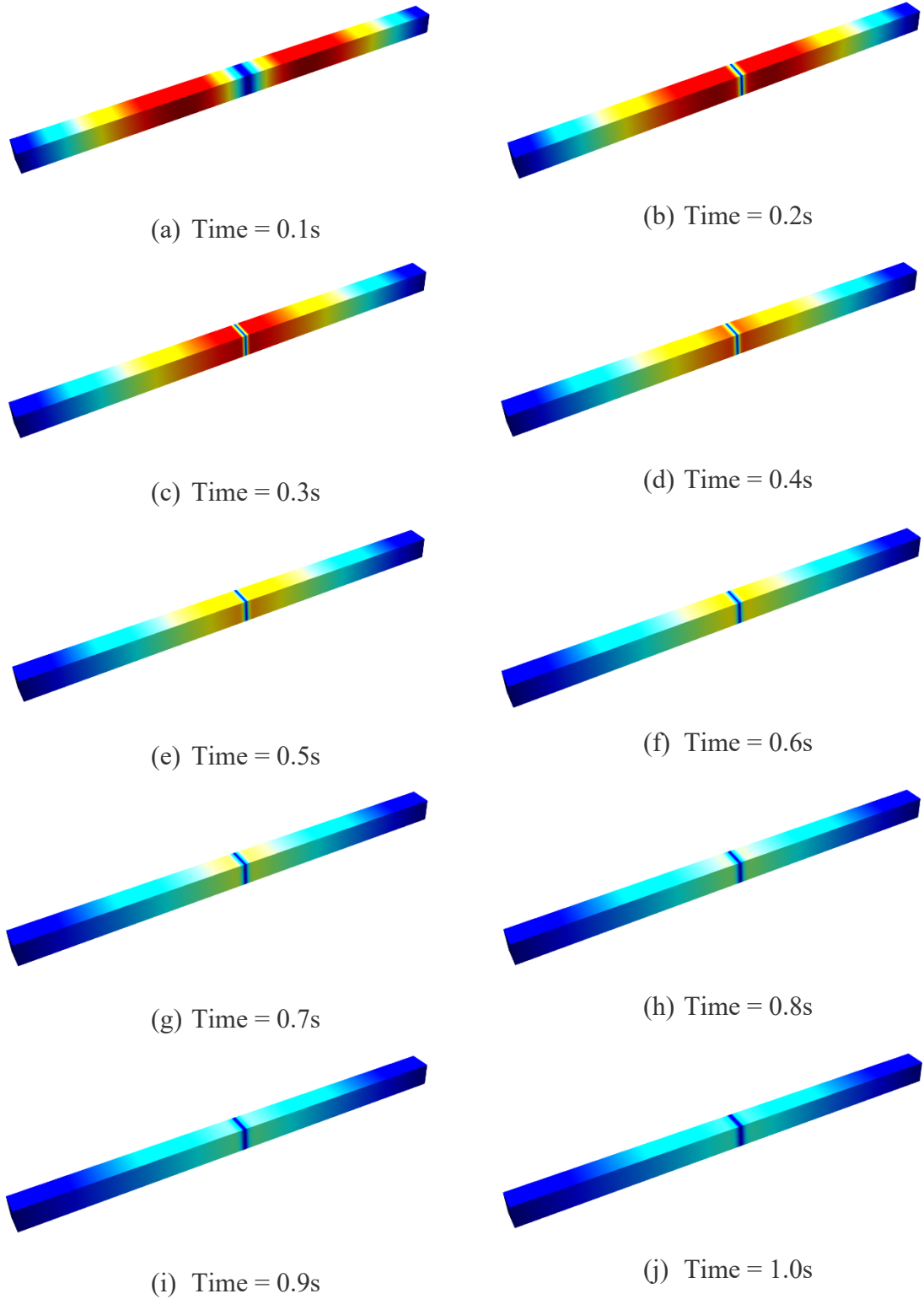

Figure 8: Predicted velocity color map at 0.1s to 1.0s (obtained via mapping predicted data back to volTypeField).

In fact, the procedure explored in this work can be generalized to a standard workflow, as presented in Figure 9 which demonstrates a streamlined workflow that integrates various mesh data formats, such as those from Fluent, StartCCM, and other industrial sources, into the OpenFOAM data structure. This process leverages OpenFOAM's robust capabilities to enhance the construction of loss functions for deep learning applications.

As illustrated, by converting third-party mesh data into OpenFOAM's format, the system can effectively manage complex geometries and apply governing equations along with boundary and initial conditions. The conversion facilitates two crucial steps in the standard workflow. The first step involves applying boundary conditions (BC) and initial conditions (IC) along with residual constraints. This is essential for correctly setting up the simulation environment and ensuring that all physical laws are appropriately integrated into the training process of the model. Following this, the model enters the training phase, where platforms like PyTorch or TensorFlow [17] are employed. This stage focuses on minimizing the loss functions, which now include physical constraints derived from the simulations, thus enhancing the model's ability to handle complex simulations and ensure physically accurate outcomes.

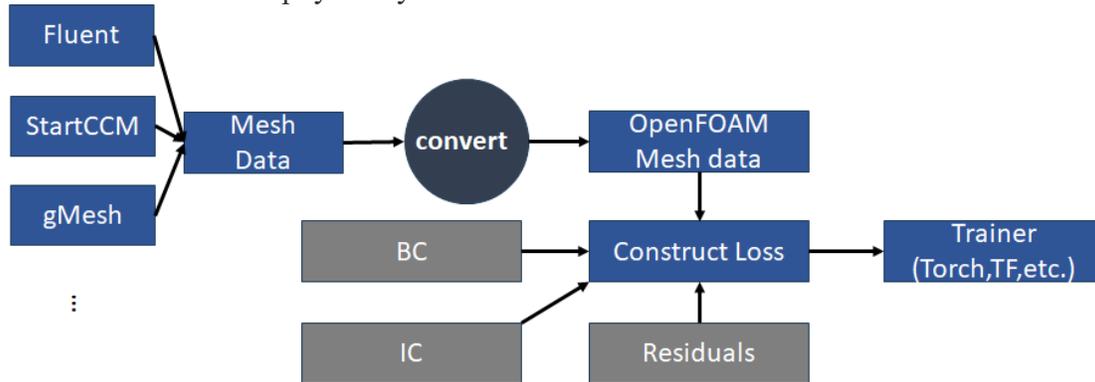

Figure 9: A possible standard workflow to utilize OpenFOAM's general data structure for Physics-informed neural network training.

However, it should be admitted that a key limitation of traditional numerical schemes is the introduction of artifacts due to discretization errors. Fully using a conventional discretization scheme will also add those artifacts. To mitigate these issues, one solution would be employing the automatic differentiation (AD) [18-20] techniques, which help reduce truncation errors by accurately evaluating the differentiation process during training. This approach not only refines the accuracy of the model but also leverages OpenFOAM's data structures to sample training points effectively across internal computational domains and complicated boundaries.

In summary, the technique exploration in the present work opens up new possibilities for employing deep learning to potentially solve intricate industrial simulation problems via leveraging robust OpenFOAM data structure.

**4 Conclusion and Future work**

This study successfully demonstrates the feasibility of integrating OpenFOAM's data structures into the training of deep learning models for computational fluid dynamics. By embedding OpenFOAM's detailed geometrical and physical constraints into neural network training, the method aligns predictions closely with traditional CFD solver outcomes. In fact, there is significant potential to apply this methodology to broader and more complex industrial challenges. Additionally, incorporating third-party mesh data into OpenFOAM's framework could extend its utility across different platforms and industrial scenarios. To further refine the approach and overcome the limitations of conventional numerical schemes, future work will explore the integration of automatic differentiation. This addition aims to enhance the accuracy and reliability of deep learning models in CFD applications, establishing a more robust framework for complex simulations